\documentclass[preprint,12pt]{elsarticle}

\usepackage{graphics}

\usepackage{amssymb}
\usepackage{latexsym}

\journal{Physica A}

\begin{document}

\begin{frontmatter}



\title{
Universal critical behavior of the
two-magnon-bound-state
mass gap for the $(2+1)$-dimensional Ising model
}


\author{Yoshihiro Nishiyama} 

\address{Department of Physics, Faculty of Science,
Okayama University, Okayama 700-8530, Japan}

\begin{abstract}

The two-magnon-bound-state mass gap $m_2$ 
for the two-dimensional quantum Ising model
was investigated
by means of the numerical diagonalization method;
the low-lying spectrum is directly accessible via the numerical diagonalization method.
It has been claimed that the ratio 
$m_2/m_1$ ($m_1$: one-magnon mass gap)
is
a universal constant
in the vicinity of the critical point.
Aiming to suppress corrections to scaling (lattice artifact),
we consider the spin-$S=1$ Ising model with finely-adjusted
extended interactions.
The simulation result for the finite-size cluster with 
$N\le20$ spins
indicates 
the mass-gap ratio $m_2/m_1=1.84(1)$.

\end{abstract}

\begin{keyword}

05.50.+q 
05.10.-a 
05.70.Jk 
64.60.-i 
\end{keyword}

\end{frontmatter}



\section{\label{section1}Introduction}

The magnons 
of the Ising ferromagnet
in the symmetry-broken phase are
attractive,
forming a bound state with a mass gap $m_2(<2m_1)$
($m_1$: one-magnon mass gap)
\cite{Zamolodchikov98,Fonseca03,Delfino04,Caselle99,Agostini97,%
Provero98,Caselle00,Fiore03,Lee01,Caselle02,%
Nishiyama08,Dusuel10}.
Recently, 
such a bound state
was observed \cite{Coldea10}
for 
a quasi-one-dimensional quantum Ising ferromagnet,
CoNb$_2$O$_6$,
by means of the inelastic neutron scattering.
A notable point is that 
the mass-gap ratio $m_2/m_1$ approaches to 
a constant value, namely,
$m_2/m_1 \to (1+\sqrt{5})/2$
(golden ratio),
 asymptotically
in the vicinity of the critical point (between the ferromagnetic
and paramagnetic phases).
As a matter of fact, 
according to the field-theoretical rigorous analysis \cite{Zamolodchikov98,Fonseca03},
the mass-gap ratio $m_2/m_1$
is a universal constant
(golden ratio) at the critical point,
providing a novel type of the critical amplitude relation;
see Ref. \cite{Delfino04} for a review.
It would be intriguing that the spectral property
is also under the reign of universality.


On the contrary, no rigorous information is available
as to
the two- (three-) dimensional quantum (classical)
Ising model, and a variety of approaches have been
made to clarify the nature of the two-magnon bound state
\cite{Caselle99,Agostini97,Provero98,Caselle00,Fiore03,Lee01,Caselle02,%
Nishiyama08,Dusuel10}; afterward, we make an overview.
The aim of this paper is to calculate
the mass-gap ratio $m_2/m_1$
for the two-dimensional spin-$S=1$ transverse-field (quantum)
Ising model
\cite{Nishiyama10}
by means of the numerical diagonalization method; 
technical details are explained in Sec. \ref{section2}.
As mentioned above,
the spin magnitude is extended to $S=1$ from $S=1/2$
\cite{Nishiyama10}.
The spin-$S=1$ model allows us to
incorporate
a variety of interactions, with which 
corrections to scaling (lattice artifact)
are suppressed considerably;
see Ref.
\cite{Hasenbusch10}, and references therein.

We make an overview of the
the preceding studies.
By means of the Monte Carlo method
for the three-dimensional 
lattice-$\phi^4$ and Ising models
up to
$  60^2\cdot120$ sites,
the mass-gap ratio was estimated
as
 $m_2/m_1=1.83(3)$ \cite{Caselle99}.
Here, the reciprocal correlation length 
$1/\xi$
was identified as the mass gap.
The result was supported by
a perturbative analysis of
the three-dimensional $\phi^4$ field theory,
$m_2/m_1=1.828(3)$ \cite{Caselle02};
the next-leading-order result, however, leads to an unphysical conclusion
$m_2/m_1<0$,
indicating a highly non-perturbative nature of this issue.
The transfer-matrix (TM) simulation for the 
three-dimensional classical Ising model 
with $N \le 15$ spins 
($N$: number of spins constituting a TM slice)
indicates
 $m_2 /m_1 =1.84(3)$ \cite{Nishiyama08}.
A recent cluster-expansion analysis of the two-dimensional
transverse-field Ising model
yields an estimate
$m_2/m_1 \approx 1.81$ \cite{Dusuel10}
through a careful resummation of the power series.

As mentioned above,
we consider the
two-dimensional spin-$S=1$ transverse-field Ising model (\ref{Hamiltonian}).
The motivation of this paper is twofold.
First, we treat a larger cluster with $N\le 20$ spins,
taking an advantage in that the Hamiltonian matrix has few
non-zero elements; note that 
a cluster with $N \le 15$ spins was simulated
\cite{Nishiyama08}
with the
transfer-matrix 
method
(for the three-dimensional classical Ising model),
where the matrix is not sparse, and  computationally demanding.
Second, 
the extension of the spin magnitude from $S=1/2$
to $S=1$
allows us to 
incorporate a variety of interactions
such as the single-ion anisotropy $D$, and the biquadratic 
exchange interactions $(J_4 , J_4')$; see Eq. (\ref{Hamiltonian}).
As mentioned afterward,
those coupling constants are finely adjusted 
\cite{Nishiyama10}
so as to eliminate
corrections to scaling, namely, the lattice artifact;
actually,
the subsequent finite-size-scaling analysis of criticality is improved 
significantly by the 
finely-adjusted coupling constants \cite{Nishiyama10}.
A recent extensive
Monte Carlo simulation for the $S=1$ Ising model
is reported in Ref. 
\cite{Hasenbusch10}.

To be specific, the Hamiltonian
for the two-dimensional spin-$S=1$ transverse-field Ising model
\cite{Nishiyama10}
is given by
\begin{eqnarray}
\label{Hamiltonian}
{\cal H}         & = & 
-J \sum_{\langle ij \rangle} S^z_i S^z_j
- J'\sum_{\langle \langle ij \rangle \rangle} S^z_i S^z_j
- J_4 \sum_{\langle ij \rangle} (S^z_i S^z_j)^2  \\
 & &
- J_4' \sum_{\langle \langle ij \rangle \rangle} (S^z_i S^z_j)^2
  +D \sum_{i} (S^z_i)^2
     -\Gamma \sum_i S^x_i
   -H  \sum_i S^z_i
              . \nonumber
\end{eqnarray}
Here, the quantum $S=1$ operators
$\{ {\bf S}_i \}$ are placed at each 
square-lattice point $i$ ($i=1,2,\dots,N$).
The summations, 
$\sum_{\langle ij \rangle}$ and
$\sum_{\langle \langle ij \rangle \rangle}$,
run over all possible nearest-neighbor and next-nearest-neighbor pairs,
respectively.
The parameters $J$ and $J'$ are the corresponding coupling constants.
(As mentioned above,
the interaction parameters, 
$D$, $J_4$, and $J_4'$, 
denote the single-ion-anisotropy, biquadratic-nearest-neighbor, 
and biquadratic-next-nearest-neighbor coupling constants,
respectively.)
The parameters $\Gamma$ and $H$ are  
the transverse and longitudinal magnetic fields, respectively.   
According to Ref. \cite{Nishiyama10},
a critical point 
locates at
\begin{eqnarray}
(J,J',J_4,J_4',D,\Gamma,H) 
&=& [
 0.4119169708 5 ,  
0.1612506961 6 ,  
-0.1176402001 8 ,    \nonumber \\  
\label{fixed_point}
& & 
-0.0 526792660 1 ,  
-0.3978195612 2   
,1.0007(17),
0
]
,
\end{eqnarray}
where corrections to scaling (lattice artifact)
are suppressed considerably.
The set of coupling constants,
Eq. (\ref{fixed_point}),
were determined through two-step procedures.
First, with $\Gamma=1$ fixed (tentatively),
the other coupling constants 
were finely adjusted to the fixed point
[Eq. (\ref{fixed_point})] of an approximate real-space
renormalization group (decimation).
(In the renormalization-group context, the irrelevant operators are almost
eliminated.)
Second, with $\Gamma$ regarded as a variable parameter,
the location of the critical point
$\Gamma=1.0007(17)$ was determined 
through the finite-size-scaling analysis (of the energy gap).
As mentioned above, right at the point (\ref{fixed_point}),
corrections to scaling are suppressed;
in the lattice-field theory,
such a lattice artifact is an obstacle to 
take the continuum limit reliably,
and the idea, the so-called perfect action, has been developed for decades
\cite{Symanzik83a,Symanzik83b,Ballesteros98,Hasenbusch99}.

\begin{figure}
\includegraphics[width=100mm]{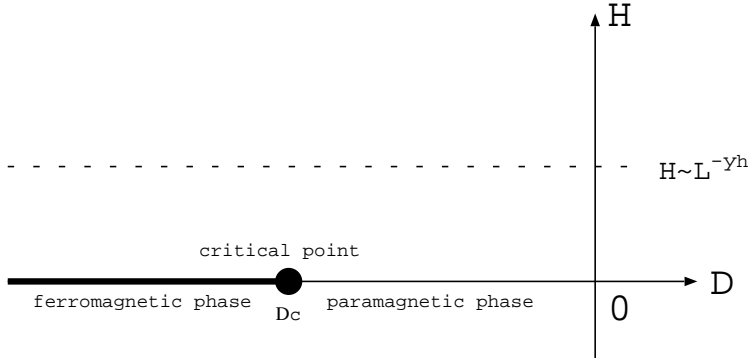}%
\caption{
\label{figure1}
The phase diagram 
of the two-dimensional spin-$S=1$ transverse-field 
Ising model (\ref{Hamiltonian})
for the single-ion anisotropy $D$
 and uniform magnetic field $H$
is presented.
[The other coupling constants are fixed to
Eq. (\ref{fixed_point}).]
The solid line denotes the first-order phase boundary,
and at the endpoint, 
$(D,H)=(D_c,0)$,
the critical point
separating
the paramagnetic and ferromagnetic phases locates.
The two-magnon bound state 
(Fig. \ref{figure2}) 
in the vicinity of the
critical point
is our concern.
The subspace of 
$H\propto L^{-y_h}$ (dashed line) 
is surveyed in Sec. \ref{section2_4}.
}
\end{figure}

The phase diagram of the model (\ref{Hamiltonian})
is presented in Fig. \ref{figure1}.
Here, the coupling constants other than $D$ and $H$ are set to 
Eq. (\ref{fixed_point}).
The $D$ term is reminiscent of the $\phi^4$-field-theory's mass term, $m^2 \phi^2$,
and the
ferromagnetic (paramagnetic) phase appears for $D<(>)D_c=-0.3978195612 2  $
($D_c$: critical point)
as anticipated.
The magnetic field is set to $H=0$ except in Sec. \ref{section2_4},
where the $H$-stabilized two-magnon bound state is surveyed.

The rest of this paper is organized as follows.
In Sec. \ref{section2}, we present the simulation results,
following an explanation of the technical preliminaries.
In Sec \ref{section3}, we address the summary and discussions.

\section{\label{section2} Numerical results}

In this section, we present the numerical results for 
the two-dimensional transverse-field
Ising model, Eq. (\ref{Hamiltonian}).
We employed
the numerical diagonalization method for the 
finite-size cluster
with $N \le 20$ spins.
We implemented the screw-boundary condition
(Novotny's method)
\cite{Novotny90}
to treat a variety of system sizes $N=8,10,\dots,20$ 
systematically;
note that conventionally, the number of spins 
is restricted within the
quadratic numbers, $N=4,9,\dots$, for a rectangular cluster.
Here, we adopt the simulation algorithm
presented in
the Appendix of Ref. \cite{Nishiyama10}.
The linear dimension $L$ of the cluster
is given by
\begin{equation}
L= \sqrt{N},
\end{equation}
because $N$ spins constitute a rectangular cluster.



\subsection{\label{section2_1}
Preliminaries: Character of
the low-lying excitation levels
}

\begin{figure}
\includegraphics[width=100mm]{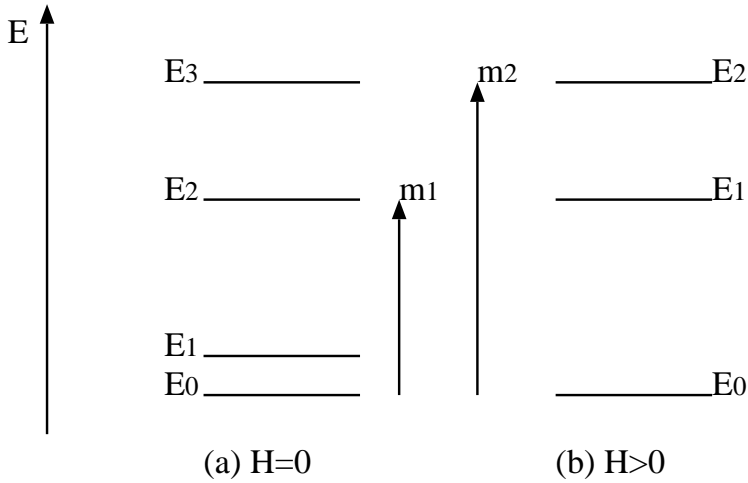}%
\caption{
\label{figure2}
A schematic drawing of the low-lying excitation levels, 
$E_0<E_1<\dots$, for the transverse-field Ising model
(\ref{Hamiltonian})
in the vicinity of the critical point
is presented.
Here, the zero-momentum subspace $k=0$ is considered;
at $k=0$, the one- and two-magnon excitation gaps,
$m_1$ and $m_2$, respectively, open.
The spectral structure depends on
(a) $H=0$, and (b) $H > 0$ significantly.
}
\end{figure}

In this section, 
we explain the technical details,
placing an emphasis on the 
character of the
low-lying spectrum; see Fig. \ref{figure2}.
As mentioned in Sec. \ref{section1},
we employed the numerical diagonalization method for the
quantum Ising model (\ref{Hamiltonian}).
The diagonalization was performed within the 
zero-momentum space,
$k=0$ (zone center),
at which 
the one- and two-magnon excitation gaps, $m_1$ and $m_2$, respectively,
open.
Hence, from the low-lying levels, $E_0<E_1<\dots$, at $k=0$,
we are able to calculate the mass gap of each excitation.
As presented in Fig. \ref{figure2},
the character of the low-lying excitation
depends significantly on 
 either (a) $H=0$ or (b) $H \ne0$.
For $H=0$, because of the quasi-degeneracy 
of the ground states,
the mass gaps $m_{1,2}$ are given by the formulas
\begin{equation}
m_1=E_2-E_0  \ , 
m_2=E_3-E_0 
.
\end{equation}
On the contrary, for $H \ne 0$,
the quasi-degeneracy becomes resolved,
and the relations
\begin{equation}
m_1=E_1-E_0 \ , 
m_2=E_2-E_0 
,
\end{equation}
hold.
It is an advantage of the numerical diagonalization method
that the low-lying levels are accessible directly.

Last, we address a technical remark.
At $H=0$, the Hamiltonian (\ref{Hamiltonian}) restores the
spin-inversion symmetry, $S^z_i \to -S^z_i$, and the
parity index characterizes the levels $E_{0,1,2,3}$;
that is,
the levels $E_{0,2}$ ($E_{1,3}$)
belong to the parity-even (odd) sector.
Practically,
the reduction of the Hilbert space with respect to
the parity index saves the computational effort,
because the evaluation of the fourth-lowest energy level $E_3$
is computationally demanding, and even unstable.


\subsection{\label{section2_2}
Finite-size-scaling analysis of $m_2/m_1$ ($H=0$)}

In this section, 
we analyze the critical behavior of the mass-gap ratio
$m_2/m_1$, 
devoting ourselves to
the subspace, $H=0$; see the phase diagram, Fig. \ref{figure1}.

To begin with, 
we consider the finite-size-scaling formula (\ref{scaling_relation}),
which sets a basis of our analysis.
(Tentatively, we turn on $H$.)
The finite-size-scaling theory insists that
the mass-gap ratio is expressed by the formula
\begin{equation}
\label{scaling_relation}
m_2 /m_1 =f \left((D-D_c)L^{1/\nu},HL^{y_h} \right)
  ,
\end{equation}
with
a certain scaling function $f$,
provided that the quantity $m_2/m_1$ is dimensionless (scale invariant)
at the critical point;
the scale invariance is confirmed by the simulation result presented
below.
The scaling parameters are set to 
the values appearing in the literatures,
$(\nu,y_h)=[ 0.63002(10)    ,2.481865(50) ]$
\cite{Hasenbusch10} and
$D_c=-0.39781956122$ \cite{Nishiyama10}.
Hence, there is no adjustable parameter 
(arbitrariness)
in the present scaling analyses.

\begin{figure}
\includegraphics[width=100mm]{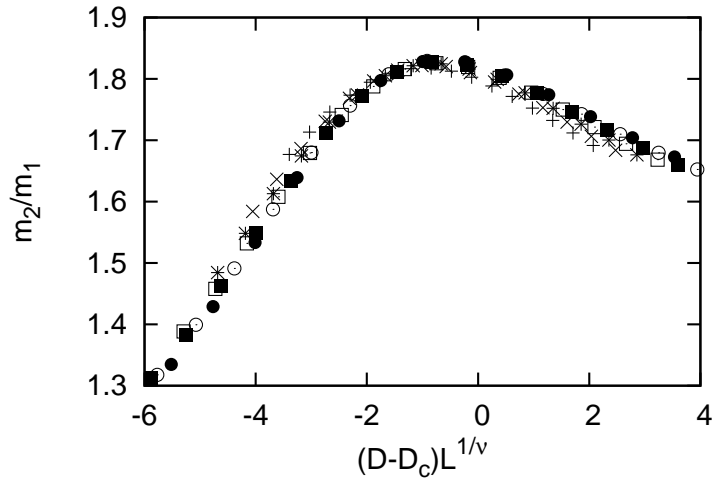}%
\caption{
\label{figure3}
The
finite-size-scaling plot,
$(D-D_c)L^{1/\nu}$-$m_2/m_1$, 
for various $D$, $N=8,10,\dots,20$, and the fixed magnetic field $H=0$
is shown;
the symbols,
$+$, $\times$, $*$, $\Box$, $\blacksquare$, $\circ$, and $\bullet$,
denote the system sizes $N=8$, $10$,
 $12$, $14$,
 $16$, $18$,
and $20$, respectively.
Here,
the scaling parameters are set to the values appearing in the literatures,
$\nu=0.63002$ \cite{Hasenbusch10} and 
$D_c=-0.3978195612 2 $ \cite{Nishiyama10}.
[The coupling constants (except $D$) are set to Eq. (\ref{fixed_point}).]
The mass-gap ratio $m_2/m_1$ 
appears to take a scale-invariant value at the critical point
$D-D_c=0$,
and the precise value is estimated in 
Fig. \ref{figure4}.
}
\end{figure}

Based on the above scaling formula
(\ref{scaling_relation}), 
we turn to the analysis of the simulation result.  
In Fig. \ref{figure3},
we present the finite-size-scaling plot,
$(D-D_c)L^{1/\nu}$-$m_2/m_1$,
for various $D$, $N=8,10,\dots,20$ and the fixed $H=0$; the other coupling constants are 
fixed
to Eq. (\ref{fixed_point}).
The data appear to collapse into a scaling curve satisfactorily;
that is,
the available system sizes already enter the scaling regime.
Moreover,
we confirm that the quantity
$m_2/m_1$ is indeed scale-invariant (dimensionless) 
at the critical point $D-D_c=0$.
Actually, the mass-gap ratio 
appears to be around
$m_2/m_1\approx 1.8$
in good agreement with the preceding estimates (see the Introduction).
In the next section, we estimate $m_2/m_1$ at the critical point,
taking the extrapolation to the thermodynamic limit.

\subsection{\label{section2_3}
The mass-gap ratio $m_2/m_1$
at the critical point}

\begin{figure}
\includegraphics[width=100mm]{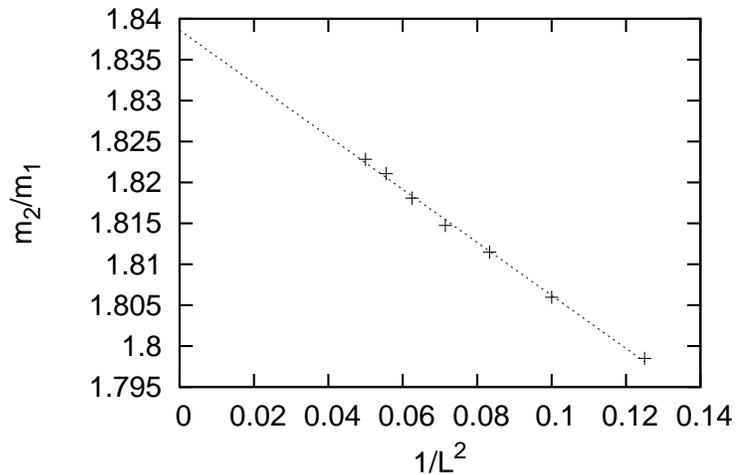}%
\caption{
\label{figure4}
The mass-gap ratio $m_2/m_1$ 
at the critical point, Eq. (\ref{fixed_point}),  is plotted for $1/L^2$
[$N(=L^2)=8,10,\dots,20$]; 
The least-squares fit to these data (dotted line)
yields 
an estimate $m_2/m_1=1.83861(62)$ in the thermodynamic limit.
The series of data exhibit a slight undulation
due to the screw-boundary condition \cite{Novotny90}.
A possible extrapolation (systematic) error
is considered in the text.
}
\end{figure}

In this section, we estimate
the mass-gap ratio $m_2/m_1$ at the critical point,
Eq. (\ref{fixed_point}).
In Fig. \ref{figure4},
we plot $m_2/m_1$ for $1/L^2$ 
[$N(=L^2)=8,10,\dots,20$] at the critical point,
Eq. (\ref{fixed_point}).  
The least-squares fit to the data 
yields an estimate $m_2/m_1=1.83861(62)$ in the thermodynamic limit.
As a reference, we made a similar least-squares-fit analysis for 
$14 \le N \le 18$, and arrived at 
a slightly enhanced estimate, $m_2/m_1=1.84311(92)$.
The discrepancy $\Delta (m_2/m_1) \approx 0.005 $
appears to dominate the least-squares-fit error $\approx 0.0006$.
As a matter of fact, the data alignment in Fig. \ref{figure4} exhibits 
a slight undulation, whose nodes locate around the system sizes of 
$N(=L^2) \approx 9,16$ (quadratic numbers).
Such an undulation is an
artifact of the screw-boundary condition \cite{Novotny90};
actually, there appears
a slight hollow around $1/L^2\approx 0.07 \sim 1/3.5^2$
(similarly, a bump around $1/L^2\approx 0.05 \sim 1/4.5^2$).
The above-mentioned discrepancy $\approx 0.005$ 
causes a systematic error, which is not appreciated properly
by the least-square-fit error.
The discrepancy seems to be bounded by,
at most, 
$1 \cdot 10^{-2}$.
Hence, we estimate the mass-gap ratio as
\begin{equation}
\label{estimate}
m_2/m_1 =1.84(1)
  .
\end{equation}
The estimate is examined by an independent analysis 
of $m_2/m_1$ ($H\ne 0$)
in the next section.

We address a number of remarks.
First, we argue the validity of the $1/L^2$-extrapolation scheme,
namely, the abscissa scale $1/L^2$ in Fig. \ref{figure4}.
In Ref. \cite{Nishiyama10}, it was demonstrated that 
the $1/L^2$-extrapolation scheme works successfully
for the analysis of the critical indices of the model concerned, Eq. (\ref{Hamiltonian});
the results turned out to be in good agreement with the
existing values.
Last, we make a comparison with the preceeding
transfer-matrix result
$m_2/m_1=1.84(3)$ 
\cite{Nishiyama08}
for $N \le 15$.
The distance between the 
extrapolated value $m_2/m_1|_{N\to\infty} =1.84$
and the raw $N=15$ result $m_2/m_1|_{N=15} \approx 1.81$
is around
$\Delta(m_2/m_1)= m_2/m_1|_{N\to\infty} -m_2/m_1|_{N=15}  \approx 0.03$.
On the one hand, the present data attain
an improved convergence
$\Delta (m_2/m_1) =  m_2/m_1|_{N\to\infty}-m_2/m_1|_{N=20}  \approx 0.017$.
Clearly, the convergence of the raw data itself is improved
possibly because of 
the expansion of the tractable system size,
and the extension of the spin magnitude to $S=1$.




\subsection{\label{section2_4}
Finite-size-scaling analysis of $m_2/m_1$ ($H \ne 0$)}

In this section, we analyze the critical behavior of $m_2/m_1$,
applying 
a properly scaled magnetic field
\begin{equation}
\label{scaled_magnetic_field}
H=A/L^{y_h}  ,
\end{equation}
with the scaling dimension of the magnetic field,
$y_h=2.481865$ \cite{Hasenbusch10}, and
a coefficient
\begin{equation}
\label{coefficient}
A=11 .
\end{equation}
Afterward, we address a remark on
the choice of $A$.
We stress that the scaled magnetic field $H=A/L^{y_h}$ vanishes
in the thermodynamic limit $L\to\infty$.
In the experiment
\cite{Coldea10},
such an infinitesimal magnetic field, the so-called
effective longitudinal mean field \cite{Carr03}, 
was induced (applied)
so as to
observe the two-magnon bound state 
beside the critical point clearly.

\begin{figure}
\includegraphics[width=100mm]{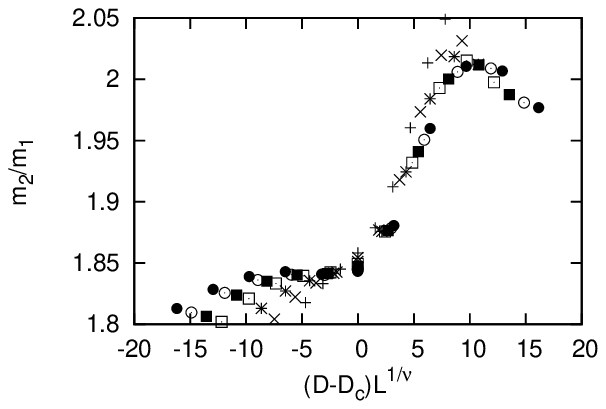}%
\caption{
\label{figure5}
The finite-size-scaling plot,
$(D-D_c)L^{1/\nu}$-$m_2/m_1$, 
for various $D$, 
$N=8,10,\dots,20$, and scaled magnetic field, $H=11/L^{y_h}$
($y_h=2.481865$ \cite{Hasenbusch10})
is shown;
the symbols,
$+$, $\times$, $*$, $\Box$, $\blacksquare$, $\circ$, and $\bullet$,
denote the system sizes $N=8$, $10$,
 $12$, $14$,
 $16$, $18$,
and $20$, respectively.
Here, the scaling parameters and the coupling constants (except $D$ and $H$) are the same as those of Fig. \ref{figure3}.
A plateau develops in the ferromagnetic phase ($D<D_c$), indicating
an existence of a $H$-stabilized bound state for a finite range of $D$.
The plateau height $m_2/m_1\approx 1.84$ supports
the 
estimate, Eq. (\ref{estimate}).
}
\end{figure}

In Fig. \ref{figure5},
we present the finite-size-scaling plot,
$(D-D_c)L^{1/\nu}$-$m_2/m_1$,
for
various $D$,
 $N=8,10,\dots,20$,
and
the scaled magnetic field $HL^{y_h}=11$,
Eq. (\ref{scaled_magnetic_field}).
The other scaling parameters and coupling constants 
(except $D$ and $H$)
are the same as those of Fig. \ref{figure3}.
The data collapse into a scaling curve, supporting
the validity of the scaling relation (\ref{scaling_relation});
in other words,
the available system sizes already enter the scaling regime.
Notably enough, in Fig. \ref{figure5},
there appears
a plateau with the height $m_2/m_1 \approx 1.84$
extending in the symmetry-broken phase, $D-D_c<0$.
Moreover, the plateau width expands gradually, as the system size enlarges.
Such a feature indicates
an existence of a two-magnon bound state with $m_2/m_1\approx 1.84$
for a wide range of the parameter space.
The observation $m_2/m_1=1.84$ supports the estimate (\ref{estimate})
obtained in Sec. \ref{section2_3}.


A number of remarks are in order.
First, we explain the choice of the coefficient $A=11$ (\ref{coefficient}).
The coefficient $A=11$ was set so as to make the plateau in Fig. \ref{figure5}
flat.
Specifically,
 the plateau slope becomes positive (negative) for $A>(<)11$.
Second, it is to be noted that the plateau height 
$m_2/m_1=1.84$
is insensitive to
the choice of
the scaling parameters such as $D_c$ and $\nu$;
these parameters simply influence the horizontal drift of the plateau,
leaving the plateau height unchanged.
In this sense, the finite-size-scaling analysis under $H\ne0$ provides unbiased 
information as to $m_2/m_1 \approx 1.84$.
Last, it has to be mentioned that in the experiment
\cite{Coldea10}, the mass-gap ratio was observed under 
a uniform magnetic field (effective longitudinal mean field \cite{Carr03}).
In the experiment,
the plateau height $m_2/m_1$ is directly observable without
carrying out the scaling analysis, because the system size
$N$ of the sample material is sufficiently large.

\section{\label{section3}Summary and discussions}

The critical behavior of the two-magnon mass gap $m_2(<2m_1)$ ($m_1$: one-magnon mass gap)
was investigated for the
two-dimensional spin-$S=1$ transverse-field Ising model
(\ref{Hamiltonian})
by means of the numerical diagonalization method; the low-lying spectrum is directly
accessible with the numerical diagonalization method, as presented in Fig. \ref{figure2}.
The universal critical behavior of
the
mass-gap ratio $m_2/m_1$ 
is our concern.
The spin-$S=1$ model (\ref{Hamiltonian})
allows us to incorporate a variety of interactions,
which are adjusted to Eq. (\ref{fixed_point}) 
so as to suppress \cite{Nishiyama10}
corrections to scaling.
As a result, we estimate the mass-gap ratio as $m_2/m_1=1.84(1)$, Eq. (\ref{estimate}).
Our result is comparable with the preceeding studies
such as the Monte Carlo result $m_2/m_1=1.83(3)$ for 
a cluster up to $60^2\cdot 120$ sites \cite{Caselle99},
the $\phi^4$-field-theoretical perturbative result
$m_2/m_1=1.828(3)$ \cite{Caselle02},
the
transfer-matrix-diagonalization result $m_2/m_1=1.84(3)$
for $N \le 15$
\cite{Nishiyama08},
and
the series-expansion result
$m_2/m_1\approx 1.81$ \cite{Dusuel10}.


It is not very clear whether a bound state exists other than $m_2$
\cite{Caselle99,Fiore03};
it is likely that either the particles with $m_1$ and $m_2$ form
a bound state with $m_3(<m_1+m_2)$ or a pair of particles $m_1$ 
constitute a
series of excited states $m_{3,4,\dots}$.
The continuum extending above $2 m_1$ prohibits us from identifying 
a single-particle-excitation branch 
out of the continuum
as in Fig. \ref{figure2}.
As a matter of fact,
for the one-dimensional quantum Ising model,
there appear
eight types of the magnon bound states with characteristic
mass gaps, $m_1<m_2<\dots<m_8$ \cite{Delfino04}, and
five branches are above the continuum threshold, $m_{4,5,\dots,8}>2m_1$.
It is suspected that the spectral function
$f( \omega )=\langle 0 | S^- ( \omega -{\cal H}+ E_0)^{-1} S^+| 0 \rangle$
with the ground-state vector $|0\rangle$ 
and the magnon-creation (annihilation) operator $S^{+(-)}=N^{-1}\sum_{i=1}^N S^{+(-)}_i$
detects a signal in
the back ground (continuum).
This problem is addressed in future study.

\section*{Acknowledgement}
This work was supported by a Grant-in-Aid 
from Monbu-Kagakusho, Japan
(Contact No. 25400402).

\bibliographystyle{elsarticle-num}

\begin{thebibliography}{00}


%
\bibitem{Zamolodchikov98}
A.B. Zamolodchikov,
Int. J. Mod. Phys. A {\bf 3} (1988) 743.


%
\bibitem{Fonseca03}
P. Fonseca and A. Zamolodchikov,
J. Stat. Phys. {\bf 110} (2003) 527.

\bibitem{Delfino04}
G. Delfino,
J. Phys. A {\bf 37} (2004) R45.






\bibitem{Caselle99} 
M. Caselle, M. Hasenbusch, and P. Provero,
Nucl. Phys. B {\bf 556} (1999) 575. 



%
\bibitem{Agostini97}
V. Agostini, G. Carlino, M. Caselle, and M. Hasenbusch,
Nucl. Phys. B {\bf 484} (1997) 331.
%
\bibitem{Provero98}   
P. Provero,
Phys. Rev. E {\bf 57} (1998) 3861.
%
\bibitem{Caselle00} 
M. Caselle, M. Hasenbusch, P. Provero, K. Zarembo,
Phys. Rev. D {\bf 62} (2000) 017901.
%
\bibitem{Fiore03} %
R. Fiore, A. Papa, and P. Provero,
Phys. Rev. D {\bf 67} (2003) 114508.

%
\bibitem{Lee01}  %
D. Lee, N. Salwen, and M. Windoloski,
Phys. Lett. B {\bf 502} (2001) 329.

\bibitem{Caselle02}
M. Caselle, M. Hasenbusch, P. Provero, and K. Zarembo,
Nucl. Phys. B {\bf 623} (2002) 474.



\bibitem{Nishiyama08}
Y. Nishiyama, Phys. Rev. E {\bf 77} (2008) 051112.

%
\bibitem{Dusuel10} 
S. Dusuel, M. Kamfor, K. P. Schmidt, R. Thomale,
and J. Vidal, Phys. Rev. B {\bf 81} (2010) 064412.


\bibitem{Coldea10}
R. Coldea, D. A. Tennant, E. M. Wheeler,
 E. Wawrzynska, D. Prabhakaran,
 M. Telling, K. Habicht, P. Smeibidl, and
K. Kiefer, Science {\bf 327} (2010) 177. 
 


\bibitem{Nishiyama10}
Y. Nishiyama,
Nucl. Phys. B {\bf 832} (2010) 605.


\bibitem{Hasenbusch10}
M. Hasenbusch, Phys. Rev. B 82 (2010) 174433.

\bibitem{Symanzik83a}
K. Symanzik, Nucl. Phys. B {\bf 226} (1983) 187.
\bibitem{Symanzik83b}
K. Symanzik, Nucl. Phys. B {\bf 226} (1983) 205.
\bibitem{Ballesteros98}
H.G. Ballesteros, L.A. Fern\'andez, V. Mart\'in-Mayor, and
A. Mu\~noz Sudupe,
Phys. Lett. B {\bf 441} (1998) 330.
\bibitem{Hasenbusch99}
M. Hasenbusch, K. Pinn, and S. Vinti,
Phys. Rev. B {\bf 59} (1999) 11471.



\bibitem{Novotny90}M.A. Novotny, J. Appl. Phys. {\bf 67} (1990) 5448.

\bibitem{Carr03}
S. T. Carr and A. M. Tsvelik,
Phys. Rev. Lett.  {\bf 90} (2003) 177206.





\end{thebibliography}







\end{document}